\documentclass[toc]{PoS}
\title{Diffraction on nuclei: effects of nucleon-nucleon correlations and inelastic shadowing within an improved Glauber-Gribov approach}

\ShortTitle{Diffraction on nuclei: effects of nucleon-nucleon
correlations and inelastic shadowing}

\author{\speaker{Claudio Ciofi degli Atti}%
         \thanks{Collaboration between INFN, Sezione di Perugia and Departamento de F\'isica,
Universidad Tecnica Federico Santa Mar\'ia  (UTFSM),  Valpara\'iso, Chile.}\\
        INFN, Sezione di Perugia, Via A. Pascoli, Perugia, Italy\\
        E-mail: \email{ciofi@pg.infn.it}}

\abstract{The importance of the effects  of nucleon-nucleon (NN) short-range
correlations (SRC) and Gribov inelastic shadowing (IS)  on various high energy scattering processes involving nuclear targets is
demonstrated within an improved Glauber-Gribov approach.}

\FullConference{Sixth International Conference on Quarks and Nuclear Physics\\
         April 16-20, 2012\\
         Ecole Polytechnique, Palaiseau,  Paris}

\newcommand{\beq}{\begin{eqnarray}}
\newcommand{\eeq}{\end{eqnarray}}

\newcommand{\be}{\begin{equation}}
\newcommand{\ee}{\end{equation}}

\def\b0{{\mbox{\boldmath$0$}}}

\def\b0{{\mbox{\boldmath$0$}}}

\def \b #1{ {\bf #1}}

\def \b #1{ {\bf #1}}

 \newcommand\beqn{\begin{eqnarray}}
 \newcommand\eeqn{\end{eqnarray}}

\def\beqy{\begin{eqnarray}}
\def\eeqy{\end{eqnarray}}

\def\Re{\,\mbox{Re}\,}

\def\sqelha{\sigma_{qel}^{hA}}
\def\sthn{\sigma_{tot}^{hN}}

\def\stota{\sigma_{tot}^{hA}}
\def\sela{\sigma_{el}^{hA}}

\def\sinha{\sigma_{in}^{hA}}

\def\eeq{\end{equation}}

\def\beqy{\begin{eqnarray}}
\def\eeqy{\end{eqnarray}}

\newcommand{\ber}{\begin{displaymath}}
\newcommand{\eer}{\end{displaymath}}
\newcommand{\bey}{\begin{eqnarray}}
\newcommand{\eey}{\end{eqnarray}}

\def\beqy{\begin{eqnarray}}
\def\eeqy{\end{eqnarray}}

\begin{document}





\section{Introduction}
It is well known that  high energy processes
involving nuclear targets, namely hadron-nucleus ($h-A$) and  nucleus-nucleus ($AA$) scattering,  provide useful
information on several phenomena like,  e.g.,  hadronization and
confinement, hadron propagation in medium, mechanisms of  formation of
high density matter, and many other ones. Most of
 theoretical approaches to describe
scattering at  multi $GeV$ energies involving nuclei,
are based upon Glauber multiple scattering theory \cite{glauber} within the
independent particle model description of the nucleus. The latter
approximation seems to be out of date, for the  nucleus is a self
bound saturated liquid  where nuclear constituents spend part of
their time in strongly correlated configurations \cite{panda}, which have been recently  experimentally
investigated in a quantitative way \cite{Subedi}. Besides nucleon-nucleon (NN) short-range
correlations (SRC), intermediate hadron-hadron
inelastic scattering (Gribov inelastic shadowing (IS) \cite{gribov}), lacking in the Glauber approach,  have to be
considered.  The importance of the effects of both SRC and Gribov IS  in several high energy scattering processes have
been
studied in a series of recent papers \cite{totalnA}-\cite{ennecoll}, finding that the two effects act frequently in the opposite directions.
The main results of these papers will be concisely illustrated in the following Sections.

\section{Formal approach}
\begin{figure}
\begin{center}
\includegraphics[width=6.5cm,height=6.0cm]{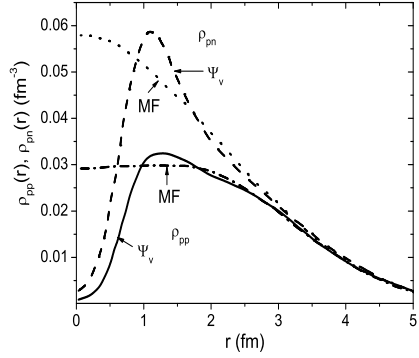}
 \vspace{-0.5cm}\caption{The proton-neutron $(pn)$ and proton-proton
$(pp)$ two-body densities $\rho_2\equiv\rho_{NN}$ in $^{16}O$ calculated \cite{panda}
within a mean field model (MF) and by solving the many-body problem
with a realistic NN interaction ($\Psi_v$). The behavior of
$\rho_{NN}$ at relative distances $r \leq 1.5\div 2\,\, fm$ is
governed by the repulsive short-range core and by the attractive
intermediate-range tensor force (after Ref. \cite{panda}). }
  \label{Fig1}
  \end{center}
\end{figure}

The nuclear quantity entering most Glauber-like calculations is
  the modulus squared of the nuclear wave function $|\psi_0|^2$,
   whose exact
expansion
 is usually truncated at  lowest order (\textit{single-density approximation})
  {\it viz} \beqn \left|\,\psi_0({\bf
r}_1,...,{\bf r}_A)\,\right|^2=\prod_{j=1}^A\,\rho_1({\bf r}_j)
\,+\,\sum_{i<j}\,\Delta({\bf r}_i,{\bf
r}_j)\hspace{-0.1cm}\prod_{k\neq i,j}\rho_1({\bf
r}_k)\,+.....\simeq \prod_{j=1}^A\,\rho_1({\bf r}_j).
\label{psiquadro}
 \eeqn
Here the \textit{two-body contraction} (\textit{two-body correlation function}) ${\Delta({\bf r}_i,{\bf
r}_j)}\,=\,\rho_2({\bf r}_i, {\bf r}_j)\,-\,\rho_{1}({\bf
r}_i)\,\rho_{1}({\bf r}_j)$, contains the effect of SRC,
represented by a hole in  the two-body density  $\rho_2$ at short
inter-nucleon separations (see Fig. \ref{Fig1}). It has been shown \cite{totalnA} that SRC   lead to an
additional contribution to the  nuclear thickness function as
follows
  (${\bf r}_i=\{{\bf s}_i,z_i\}$) \cite{totalnA}
\be
  \Delta T_A^h(b)=
  \frac{1}{\sthn}
  \int d^2{\bf s}_1\,d^2{\bf s}_2\,
  \Delta^\perp_A({\bf s}_1,{\bf s}_2)
  \Re \Gamma^{pN}({\bf b}-{\bf s}_1)\,
  \Re \Gamma^{pN}({\bf b}-{\bf s}_2),
  \label{55}
  \eeq
  \begin{figure}
\begin{center}
\vspace{-0.5cm}
\includegraphics[width=7.0cm,height=6.0cm]{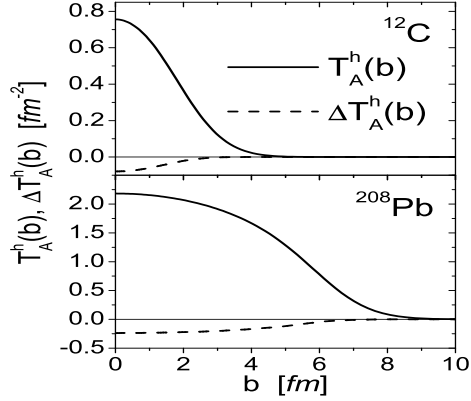}
\caption{The thickness function $T_A^h(b)$ and the correlation
contribution ($\Delta T_A^h(b)$)   in $p-^{12}C$ and $p-^{208}Pb$
collisions  at HERA-B energies. The total thickness function is
given by ${\widetilde T}_A^h= T_A^h- \Delta T_A^h$
(after Ref.\cite{nashboris})}
\label{Fig2}
  \vspace{-0.5cm}
  \end{center}
  \end{figure}
\noindent where $\Delta^\perp_A({\bf s}_1,{\bf s}_2)$
  represents the transverse two-nucleon contraction and the total thickness function
is ${\widetilde T}_A^h= T_A^h- \Delta T_A^h$. The thickness
functions of $^{12}C$ and $^{208}Pb$ at HERA-B energies calculated with realistic two-body densities
from Ref. \cite{ACMprl}
 are shown
in Fig. \ref{Fig2}. SRC increase the thickness functions and,
consequently, the total neutron-nucleus cross section at high
energies, making the nucleus more opaque \cite{totalnA},
 unlike  Gribov IS corrections
  which increase nuclear transparency.
  \begin{figure}
\vskip 0.5cm
  \begin{center}
\includegraphics[width=10.8cm,height=11.5cm]{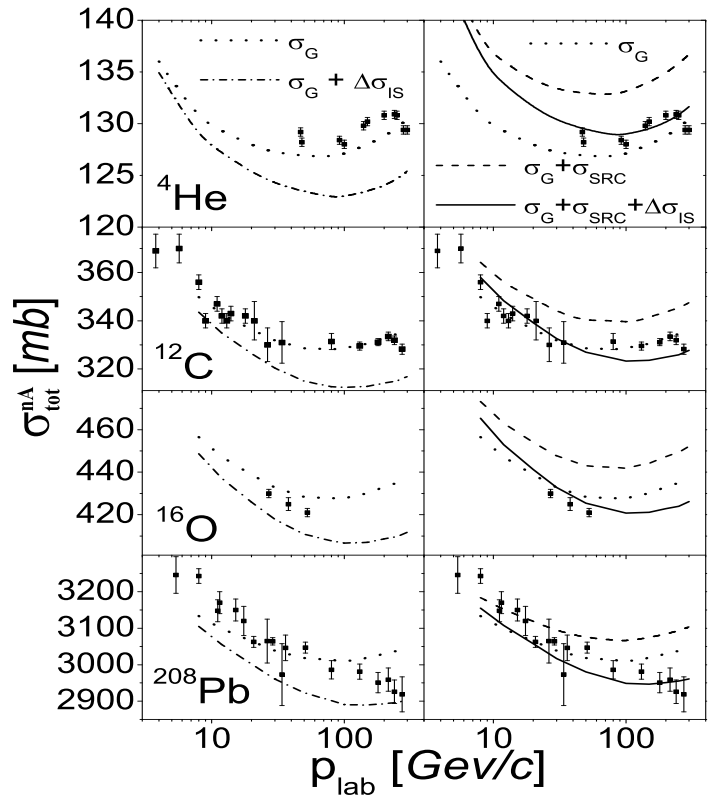}
\caption{The total neutron-Nucleus cross section $\sigma_{tot}^{nA}$ {\it vs} $p_{lab}$. \textit{Left panel}:
  Glauber single density approximation ($\sigma_G$; \textit{dots})
  and Glauber plus Gribov inelastic shadowing
  ($\sigma_G$ + $\Delta\sigma_{IS}$; \textit{dot-dash}).
  \textit{Right panel}:
  Glauber ($\sigma_G$; \textit{dots});
  Glauber   plus SRC ($\sigma_G$ + $\sigma_{SRC}$; \textit{dashes});
  Glauber   plus SRC
  plus Gribov inelastic shadowing  ($\sigma_G$ + $\sigma_{SRC}$ + $\Delta\sigma_{IS}$;
    \textit{full})
  (after Ref. \cite{totalnA}).}
\label{Fig3}
\end{center}
\end{figure}
\section{Results of calculations}
\begin{table}[!htp]
\begin{center}
 \begin{tabular*}{0.78\textwidth}{@{\extracolsep{\fill}}c| c c c c }\hline\hline
$^{208}Pb$& Glauber & Glauber & q-2q model & 3q model \\
& & +SRC & +SRC & +SRC \\\hline $\sigma_{tot}^{NA}\,\,\,  [mb]$ &
3850.63 & 3885.77 & 3833.26 & 3839.26 \\\hline $\sigma_{el}^{NA}
\,\,\,[mb]$  & 1664.76 & 1690.48 & 1655.70 & 1660.67 \\\hline
$\sigma_{sd}^{NA}\,\,\, [mb]$ & - & - & 2.62 &    0.59 \\\hline
$\sigma_{sd+g}^{NA} [mb]$& - & - & 2.58 & 2.56
\\\hline $\sigma_{qe}^{NA}\,\, \, [mb]$  &  120.92 &  112.65 & 113.37 & 113.88
\\\hline $\sigma_{qsd}^{NA}\,\,\, [mb]$ & -       & -       & -2.08 & -2.62
\\\hline $\sigma_{tsd}^{NA}\,\,\, [mb]$ & -       & -       & 17.55 & 17.63
\\\hline $\sigma_{dd}^{NA}\,\, \, [mb]$  & -       & -       & -2.08 & -2.62
\\\hline\hline
\end{tabular*}
 \caption{Various $p-^{208}Pb$ cross sections at LHC energies  calculated with two models of the light-cone dipole cross section
 (after Ref. \cite{nashboris}).}
 \end{center}
\end{table}
An exhaustive calculation
 of the total, $\stota$, elastic, $\sela$,
quasi-elastic, $\sqelha$, inelastic, $\sinha$, and diffractive
dissociation hadron-nucleus cross sections, which include both SRC
and Gribov IS summed to all orders by the light-cone dipole approach
\cite{KLZ,boris1,boris2}, has been performed  in Ref. \cite{totalnA,nashboris,ennecoll} demonstrating  the opposite roles played by
SRC and IS.
 It has been found that the total contribution to the thickness function due to SRC and Gribov IS reads as follows
 \begin{eqnarray}
  &&\hspace{-0.5cm}\Delta T_A^{dip}( b,{\bf r}_T,\alpha)=\nonumber\\
  &&\hspace{-0.5cm}=\frac{1}{\sigma_{dip}(r_T)}
  \int d^2{\bf s}_1\,d^2{\bf s}_2\,
  \Delta^\perp_A({\bf s}_1,{\bf s}_2)
  \Re \Gamma^{{\bar q} q,N}({\bf b}-{\bf s}_1,{\bf r}_T,\alpha)
  \Re \Gamma^{{\bar q} q,N}({\bf b}-{\bf s}_2,{\bf r}_T,\alpha).
  \label{500}
  \end{eqnarray}
where $\Gamma^{q{\bar q},N}$ is the $(q{\bar q})-N$ profile, $\sigma_{dip}(r_T)$ the dipole cross section, $r_T$ the dipole transverse
separation and
$\alpha$ its fractional light-cone momentum. SRC and Gribov IS affect also the number of inelastic
collisions $N_{coll}^{hA}=A\,\sigma_{in}^{hN}/\sigma_{in}^{hA}$,  which is the normalization
factor  used to obtain the nucleus to
nucleon ratio of the cross section of a hard reaction. The results
of  calculations, performed with  realistic one- and two-body
densities and correlation functions from Ref. \cite{ACMprl}, are shown
in Fig. \ref{Fig3},  and in Tables 1 and 2.
\begin{table}[!h]
\begin{tabular*} {0.98 \textwidth}{@{\extracolsep{\fill}}c| c
c c c c c c}\hline\hline & & & GLAUBER & & & \\ \hline &
$\sigma_{in}^{NN}\:[mb]$ & $\sigma_{tot}^{NA}\:[mb]$ &
$\sigma_{el}^{NA}\:[mb]$ & $\sigma_{qel}^{NA}
\:[mb]$ & $\sigma_{in}^{NA}\:[mb]$& $N_{coll}$\\
\hline RHIC &42.10  &3297.56 &1368.36 &66.06 &1863.14 &4.70 \\
\hline LHC  &68.30  & 3850.63 &1664.76 &120.92 &2064.95 &6.88\\
\hline \hline & & & GLAUBER+SRC & & & \\ \hline &
$\sigma_{in}^{NN}\:[mb]$ & $\sigma_{tot}^{NA}\:[mb]$ &
$\sigma_{el}^{NA}\:[mb]$ &
$\sigma_{qel}^{NA}\:[mb]$ & $\sigma_{in}^{NA}\:[mb]$  & $N_{coll}$\\
\hline RHIC &42.10  &3337.57 &1398.08 &58.47 &1881.02 &4.65\\
\hline LHC  &68.30  & 3885.77& 1690.48&112.65 & 2082.64&6.82\\
\hline \hline & & & GLAUBER+SRC+GRIBOV($q-2q$)& & & \\ \hline
& $\sigma_{in}^{NN}\:[mb]$ & $\sigma_{tot}^{NA}\:[mb]$ & $\sigma_{el}^{NA}\:[mb]$ & $\sigma_{qel}^{NA}\:[mb]$ & $\sigma_{in}^{NA}\:[mb]$  & $N_{coll}$\\
\hline RHIC &42.10  &3228.11 &1314.04 &71.99 & 1842.08 &4.75 \\
\hline LHC  &68.30  &3833.26 &1655.70 &113.37 & 2064.19 &6.88 \\
\hline \hline
 \end{tabular*}
 \caption{Number of inelastic collisions $N_{coll}$  in $p-^{208}Pb$ scattering at
  RHIC and LHC energies (after Ref. \cite{ennecoll}).}
\end{table}
The behavior of $N_{coll}^{NA}$ is entirely governed by the
non-diffractive cross section $\sigma_{in}^{NA}$ which, as shown in Table 2, is
decreased by SRC and increased by Gribov IS. The effects of both
SRC and Gribov IS amount to few percent in agreement with the
results of the calculation of deuteron-gold scattering
\cite{boris1}. A further step in this type of calculations is to consider  the high energy collision of two
nuclei, $A$ and $B$; in this case   the correlation contribution to the
thickness function can be written as follows
\begin{figure}
\begin{center}
\vspace{-0.5 cm}
\includegraphics[width=7.5cm,height=5.0cm]{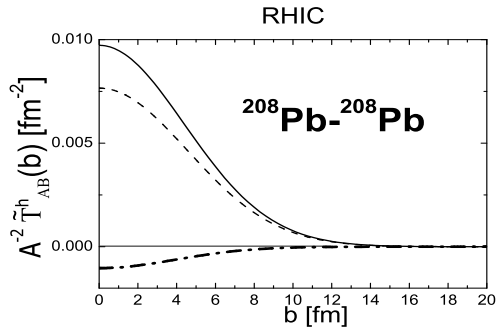}
\vspace{-0.3cm} \caption{Uncorrelated thickness function
$T_{AB}^h(b)/A^2$ (dash);  correlation contribution $\Delta
T_{AB}^h(b)/A^2$  (dot-dash);  total thickness function
  $ {\widetilde T}_{AB}^h/A^2 = [T_{AB}^h(b) -2 \Delta T_{AB}^h(b)]/A^2$ (full)
   in $^{208}Pb
-^{208}Pb$ collisions at RHIC energies}
  \label{Fig4}
  \end{center}
\end{figure}
\beqn
&&{\Delta{T}}^h_{AB}(b)= \frac{1}{\sigma^{NN}_{tot}}\,A_A\,A_B^2 \times\nonumber\\
&&\times\int d^2\,s_{A} \rho _A({\bf s}_A) \int d^2s_{B1}
d^2s_{B2} \Delta^{\perp}_B({\bf s}_{B1},{\bf s}_{B2})
\Gamma^{NN}({\bf  b} - {\bf s}_A + {\bf s}_{B1})\Gamma^{NN}({\bf b}
- {\bf
s}_A + {\bf s}_{B2}) +\nonumber\\
&&+\{ A\longleftrightarrow B\}
 \label{Nucleus_fin2}
 \eeqn
 where the 1st term  describes the interaction of a nucleon in $A$ with
 two correlated nucleons in $B$ and the 2nd term in figure brackets viceversa.
The thickness function including the effects of SRC in
$^{208}Pb-^{208}Pb$ scattering at RHIC energies is shown in Fig.
\ref{Fig3}. Calculations of possible effects of SRC in heavy ion scattering are in progress.
\section{Acknowledgments}
I am grateful to Boris Kopeliovich, Irina
Potashnikova,  Ivan Schmidt, UTFSM, Massimiliano Alvioli and Chiara Benedetta Mezzetti, INFN,  for a stimulating collaboration.


\end{document}